\documentclass[prl,aps,preprint,showkeys]{revtex4}
\begin{document}
\setcounter{page}{1}
\date{\today}
\title{A Random Multifractal Tilling}

\author{  M. G. Pereira,  G. Corso$\dagger$,  L. S. Lucena, and J. E. Freitas } 

\affiliation{ International Center for Complex Systems and    
 Departamento de F{\'\i}sica Te\'orica e Experimental,      
 Universidade Federal do Rio Grande do Norte, Campus Universit\'ario   
 59078 970, Natal, RN, Brazil.}

\affiliation{ $\dagger$ Departamento de Biof{\'\i}sica,
    Centro de Bioci\^encias,
Universidade Federal do Rio Grande do Norte, Campus Universit\'ario
 59072 970, Natal, RN, Brazil.}

\begin{abstract}
We develop a multifractal random tilling that fills the square. The multifractal 
is formed by an arrangement of rectangular blocks of different 
sizes, areas and number of neighbors. The overall feature of the tilling is 
an heterogeneous and anisotropic random self-affine object. 
 The  multifractal is constructed by an algorithm that makes successive sections of 
the square. At each $n$-step there is a random choice of a parameter $\rho_i$ related 
to the section ratio. For the 
case of random choice between $\rho_1$ and $\rho_2$ we find analytically the full 
spectrum of fractal dimensions. 

\end{abstract}


\keywords{  multifractal, fractal spectrum, random object, 
complex systems}

\maketitle

\section{I - Introduction}

In the reference \cite{last,lim} a deterministic 
multifractal tilling
is developed to study percolation in multifractal systems.
This object is a natural generalization of the square lattice in
which it is possible to estimate analytically the full spectrum
of fractal dimensions. Besides it exhibits a rich distribution
of area among the blocks and a non-trivial topology, the
number of neighbors varies along the object.   
The multifractal, however, is deterministic in its construction. In 
spite of its properties it lacks the random ingredients found in natural 
systems. In order to make the model more realistic we develop 
a random version of this model. 

In this work we develop a multifractal that has the following 
characteristics. It has an infinite number of subsets each one with 
a distinct fractal dimension. For a specific case we 
determine analytically the dimension of each 
subset. The sum of all the subsets fills completely the square. 
The algorithm of construction of the multifractal has two or more 
free parameters, $\rho_i$, that are related to the stretching of its subsets. 
During the construction of the multifractal there is random choice of 
$\rho_i$ which makes the object non deterministic. 
For the special case of just one parameter $\rho=1$ the object 
degenerates into the square lattice. The extension of the model 
for an arbitrary dimension is trivial. Finally the algorithm for 
construction of the multifractal is simple and easily implemented 
in the computer.

 Geologic structures can show multifractal 
characteristics \cite{Muller1,Muller2,Riedi,Hubert,Lovejoy} in  
physical properties like porosity, permeability, or sound 
velocity. In addition the rocks and strata are not deterministic 
objects, but present random features. 
These geologic systems  that combine multifractality and randomness  
 are natural candidates to be modeled by the proposed 
model.  

The deterministic multifractal 
is an intuitive generalization of the square lattice. Suppose that in the 
construction of the square 
lattice we use the following algorithm: take a square  
of size $1$ and 
cut it symmetrically with  vertical and  horizontal lines. 
Repeat this process 
$l$-times; at the $l^{th}$ step we have a regular lattice 
with $2^l \times 2^l$ 
cells. The setup algorithm of the deterministic multifractal is quite similar, 
the main difference 
is that we do not  cut the square in a symmetric way, but according 
to a given constant ratio. 
The random multifractal is an extension of the deterministic model. In the 
random multifractal  we use a non constant ratio in the algorithm setup. 

The main objective of this work is to present a random multifractal 
tilling and determine its set of fractal dimensions for a specific case. 
The manuscript is organized as follows: 
 in section $2$ we present the deterministic multifractal and 
review the derivation of its  
set of fractal dimensions.  In section $3$ we develop the 
random multifractal and indicate a formula that  estimate 
its set of fractal dimensions for the case of random choice between 
two given distinct $\rho_i$. 
 Finally in section $4$ we summarize the main differences
between deterministic and random multifractal tillings and give 
our final remarks.

\section{II - The  Deterministic Multifractal}

 We start with a square of linear size $1$ and a  parameter
$0<\rho = \frac{s}{r} <1$, for integers $r$ and $s$. 
The first step, $n=1$, consists of making one vertical section of 
the square.  The square is cut in two pieces of 
areas $\frac{s}{r+s}=\frac{\rho}{1+\rho}$ 
and $\frac{r}{s+r}=\frac{1}{1+\rho}$ by a vertical segment. 
This process is shown in figure \ref{fig1} (a), where we use
$\rho =  \frac{2}{3}$.
 The step $n=2$ consists of a horizontal cut with the same 
section $\rho$, this step is shown in figure \ref{fig1} (b).
At this point there are 
four rectangular blocks: the largest one of area $(\frac{1}{1+\rho})^2$,
two  of area $\frac{\rho}{(1+\rho)^2}$ and the
smallest one of area $(\frac{\rho}{1+\rho})^2$. 
 We call a set of all elements with the same area as 
a $k$-set. At the step $n=2$ there are three $k$-sets.

In the third step, $n=3$, we
repeat the same process of vertical 
section as in step $n=1$. At $n=4$ we make 
 new horizontal sections.  All the 
sections have the same ratio $\rho$.  
At the $n^{th}$-step there are $2^{ n}$ blocks.
The partition process produces a set of blocks with a variety of areas.
At the step $n$ we have $n+1$ $k$-sets. The construction process of 
the deterministic multifractal is done by vertical and horizontal sections 
of the square. As the section ratio is always the same and the segment section is 
 displaced in space the object shows  self-affinity.

 At the $n^{th}$-step of the algorithm, the 
partition of the area of the square 
 follows the binomial rule:
\begin{equation}
    1 \> = \> \sum_{k=0}^n \> C_k^n \> \left(\frac{s}{s+r}\right)^k 
          \left(\frac{r}{s+r}\right)^{n-k} 
                = \> \left(\frac{r+s}{r+s}\right)^n,
\label{bino1}
\end{equation}
or:
\begin{equation}
    1 \> = \sum_{k=0}^n C_k^n \left(\frac{\rho}{1+\rho}\right)^k 
\left(\frac{1}{1+\rho}\right)^{n-k}
= \left(\frac{1+\rho}{1+\rho}\right)^n.
\label{bino2}
\end{equation}
The number of elements of each $k$-set in this case is $C_k^n$.

Each $k$-set determines a specific dimension. 
For the estimation of  the spectrum  of fractal dimension 
$D_k$ of an object $X$ we use the box
counting method \cite{barnsley}. The object $X$ is
immersed in the plane of real numbers
 $\Re^2$, we use the trivial metric. We cover the object 
  by just-touching square boxes
of side length $\epsilon$. Let $N(X)$ denote the number of square cells
 of side length $\epsilon$
which intersect $X$. If
\begin{equation}
                 D_X = lim_{\epsilon \rightarrow 0}
          \frac{log \> N(X)}{log \> \frac{1}{\epsilon}},
\label{boxcou1}
\end{equation}
 is finite, then $D_X$ is the dimension of the $X$.

 In our case the object $X$ is a $k$-set. Remember that
the $k$-set corresponds to a set of rectangles of
the same area. If the square is
partitioned $n$ times  we have $\epsilon = 1/(s+r)^{n}$. 
For each $k$-set the area of blocks (using $\epsilon$ area units) is done by: 
\begin{equation}
                 N_k = C_n^k \> \> s^k \>r^{(n-k)},
\label{NNN}
\end{equation}
where $C_n^k$ is the binomial coefficient that express
the number of elements $k$-type, and
$s^k \> r^{(n-k)}$ is the area of each element of this set. 
 We put together all these elements to have  the fractal dimension of
each $k$-set:
\begin{equation}
                 D_k = lim_{\epsilon \rightarrow 0} \frac{ log N_k }{ log \frac{1}{\epsilon }} =
 lim_{n \rightarrow \infty} \frac{log \> C_n^k \>
\> s^k \>r^{(n-k)} }{log \> (s+r)^n }.
\label{boxcou2}
\end{equation} 
The case  $r=s=1$ is degenerated. In this situation the 
 subsets of the lattice are composed uniquely by square cells of the same area. 
Therefore the tilling is formed by a single subset of dimension $2$.

Figure \ref{fig2} shows the spectrum of $D_k$ for several values of $\rho$. 
We plot in the $x$-axis the normalized $k$-set, $k/n$, and in the $y$-axis 
the respective $D_k$. The dashed line corresponds to a $\rho$ that is very 
closed to $1$ in this situation the blocks have almost the same area and 
$D_k$ is a symmetric curve. In the opposite situation 
$\rho \rightarrow 0$ we have curves that 
correspond to a tilling in which the 
blocks are very stretched in one direction. In the figure it is also depicted 
the curves corresponding to $\rho = 0.4$, $0.6$ and $0.8$.

\section{III - The Random Multifractal}

In the last section we use a constant $\rho$ through  
all the $n$-steps of the construction of the 
multifractal. A more realistic generalization of the above process is 
to use different values of the ratio $\rho$ along the 
growing process of the multifractal. In the general case a 
random $\rho_i$ is used at each step of the algorithm. 
In this work we explore in more detail the case where at each $n$-step of the 
construction of the multifractal 
we randomly choose between two rational 
values $\rho_{1}=s_1/r_1$ and $\rho_{2}=s_2/r_2$ for $r_1+s_1=r_2+s_2$. 
The picture of this process 
is similar to figure \ref{fig1}, but this time the section 
of the square does not follow a constant ratio making the partition 
of the square asymmetric.

 We have in the algorithm of the random 
multifractal the following choices. At $n=1$ we divide the 
square in two blocks. There are two partition possibilities 
for the square:  
$\rho_{1}/(1+\rho_1)$ and $(1-\rho_{1})/(1+\rho_1)$ or $\rho_{2}/(1+\rho_2)$ e 
$(1-\rho_{2})/(1+\rho_2)$. 
At the second step, $n=2$, there are  two possibilities for 
each former configuration. These are the three partition possibilities 
 at step $n=2$:

$\rho_{1}^{2}/(1+\rho_1)^2$, $\rho_{1}(1-\rho_{1})/(1+\rho_1)^2$, 
$(1-\rho_{1})^{2}/(1+\rho_1)^2$,
$\rho_{1}(1-\rho_{1})/(1+\rho_1)^2$,

 or $\rho_{1}\rho_{2}/(1+\rho_1)(1+\rho_2)$,
$(1-\rho_{1})\rho_{2}/(1+\rho_1)(1+\rho_2)$, $\rho_{1}(1-\rho_{2})/(1+\rho_1)(1+\rho_2)$,
$(1-\rho_{1})(1-\rho_{2})/(1+\rho_1)(1+\rho_2)$,

 or $\rho_{2}^{2}/(1+\rho_2)^2$,
$\rho_{2}(1-\rho_{2})/(1+\rho_2)^2$, $(1-\rho_{2})^{2}/(1+\rho_2)^2$,
$\rho_{2}(1-\rho_{2})/(1+\rho_2)^2$. 

Note that there are not four configurations  
but only three. In fact, it is the same to make 
a section $\rho_{1}$ after $\rho_{2}$, or 
$\rho_{2}$ after $\rho_{1}$.

At the step $n$, $\rho_{1}$ is chosen $l$ times and 
$\rho_{2}$ is chosen $m$ times for $l+m=n$. At each step the 
areas of the blocks are obtained by the multiplication of the area of the block 
of the former step. Therefore the partition of 
the area  of the square is done by the following equation: 
$$
    1=\sum^{l}_{i=0}{\sum^{m}_{j=0}{C^{l}_{i}
\left(\frac{\rho_{1}}{1+\rho_1}\right)^{i}
\left(\frac{1-\rho_{1}}{1+\rho_1}\right)^{l-i}\> C^{m}_{j}
\left(\frac{\rho_{2}}{1+\rho_2}\right)^{j}
\left(\frac{1-\rho_{2}}{1+\rho_2}\right)^{m-j}}}
$$
\begin{equation}
    =\left(\frac{1+\rho_{1}}{1+\rho_1}\right)^{l}
\left(\frac{1+\rho_{2}}{1+\rho_1}\right)^{m}.
\label{NN1}
\end{equation}
This formula is analogous to equation (\ref{bino2}) of 
the deterministic case. The occurrence of blocks 
of the same $k$-sets (objects of same 
area) is smaller for this case compared with the deterministic model. 
Otherwise, the quantity of $k$-sets is larger, it is done by 
$(l+1)(m+1)$. 

The determination of the spectrum of fractal dimensions for the 
case of the multifractal is similar to the 
deterministic case, equation (\ref{boxcou2}).  
The equation for $D_k$ in this case is the following:
\begin{equation}
                 D_k = lim_{n \rightarrow \infty} \frac{log \> C^{l}_{i} s_{1}^{i}
r_{1}^{l-i} \> C^{m}_{j} s_{2}^{j} r_{2}^{m-j}  }
                      {log \> (s_1+r_1)^n },
\label{random}
\end{equation}
where the set of $k$ corresponds to any combination of $l$ and $m$ such that $n=l+m$.  
The generalization of this formula for several $\rho_i$ (for the case 
where each $\rho_i$ obeys $r_i+s_i=r_1+s_1$) is performed 
by the introduction of new terms  
$C^{m_i}_{j} s_{2}^{j} r_{2}^{m_i-j}$ in the numerator.

Figure \ref{fig3} shows the curve in 
crescent order of fractal dimensions $D_k$. 
In the $y$-axis is plotted $D_k$ and in the $x$-axis, $(k/n)^\star$, 
the respective  $k$-set.  
We notice that the 
$k$-sets in the $x$-axis are shown in order of crescent $D_k$. 
The area of the $k$-sets is normalized between zero and one. 
In the figure there are three curves $\rho_1=1/4$ (point line), 
$\rho_2=2/3$ (dashed line), and the random curve (thick line).  
In this  simulation we use $n=100$. The random curve is defined as the spectrum 
that results from the random choice between $\rho_i$ and $\rho_2$. 
The estimation of the random curve is 
performed assuming that for $n \rightarrow \infty$ we have $m=j=n/2$.

A typical view of the the deterministic and the random multifractals 
is shown in figure \ref{fig4}. These two 
pictures show the deterministic multifractal (a) and one 
realization of the random multifractal (b). In figure \ref{fig4} (a) we use 
$\rho=1/5$ and in figure \ref{fig4} (b) $\rho_1=1/4$ and $\rho_2=2/3$. In both 
situations we evolved the algorithm until $n=10$.

\section{IV - Final Remarks}

In this work we develop a random self-affine multifractal object that forms a tilling 
 of the square.  
The  multifractal is constructed by an algorithm 
that make successive sections of 
the square. In the special case where the section ratio is $1/2$ the algorithm 
generates the well known square lattice.   At each $n$-step there 
is a random choice of partition parameter $\rho_i$. 
We explore in detail the case in which the ratio choice done is between 
two specific values $\rho_1$ and $\rho_2$. For this situation  
at the limit of infinite partition of the square, $n \rightarrow \infty$, we 
find analytically  the full spectrum of fractal dimensions.

In the references \cite{last,lim,ulti} the deterministic multifractal object 
was used to study percolation properties and the breaking of 
universality class in two dimensions. In fact,  the  critical 
exponent related to the correlation length for the deterministic multifractal
 is not universal, it means, 
it is not the same as in standard percolation in two dimensions. This fact 
is related with the breaking of local isotropy. 
There are some points to consider in the
symmetry properties of this multifractal object and its isotropy.
The first is related with the stretching of
the blocks of the multifractal which increases as $\rho \rightarrow 0$. Moreover
 the topology of multifractal, specially its coordination number,  
changes along the object.  
The  object we develop 
in this work  is a natural candidate to extend the 
cited papers to a non deterministic 
analyses of the breaking of universality class. 
In special we intend to analyze the reason for the breaking of universality class  
in the random multifractal and how the local isotropy affects this process.

\vspace{2cm}

The authors gratefully acknowledge the financial support of Conselho Nacional
de Desenvolvimento Cient{\'\i}fico e Tecnol{\'o}gico (CNPq)-Brazil, 
FINEP and CTPETRO.

\hspace{1cm}

\centerline{FIGURE LEGENDS}

\begin{figure}[ht]
\begin{center}
\caption{ The initial two steps in the formation of 
the deterministic multifractal. 
In (a) a vertical line cut the square in two pieces according to  $\rho$.
Two horizontal lines sectioning the rectangles by
 the same ratio are depicted in (b). The same process is 
repeated $n$-times inside each block. At the limit $n \rightarrow \infty$ 
we have a multifractal. }
\label{fig1}
\end{center}
\end{figure}
\begin{figure}[ht]
\begin{center}
\caption{ The spectrum of fractal dimensions, $D_k$, for the deterministic multifractal.
In the $y$-axis is $D_k$ and in the $x$-axis the respective normalized $k$-set.
The curves for several $\rho$ are indicated in the figure. 
}
\label{fig2}
\end{center}
\end{figure}
\begin{figure}[ht]
\begin{center}
\caption{ The spectrum $D_k$ for $\rho_1=1/4$ (point line) 
and $\rho_2=2/3$ (dashed line). In thick line is shown the random curve. 
In the  the $x$-axis we show $(k/n)^\star$ the $k$-sets 
in order of crescent $D_k$.  
The area of the $k$-sets  are normalized to one. }
\label{fig3}
\end{center}
\end{figure}
\begin{figure}[ht]
\begin{center}
\caption{ The random multifractal 
tilling evolved until $n=10$ for (a) the deterministic 
multifractal ($\rho=1/3$) and (b) one realization of the random multifractal 
(for $\rho_1=1/4$ and $\rho_2=2/3$).}
\label{fig4}
\end{center}
\end{figure}

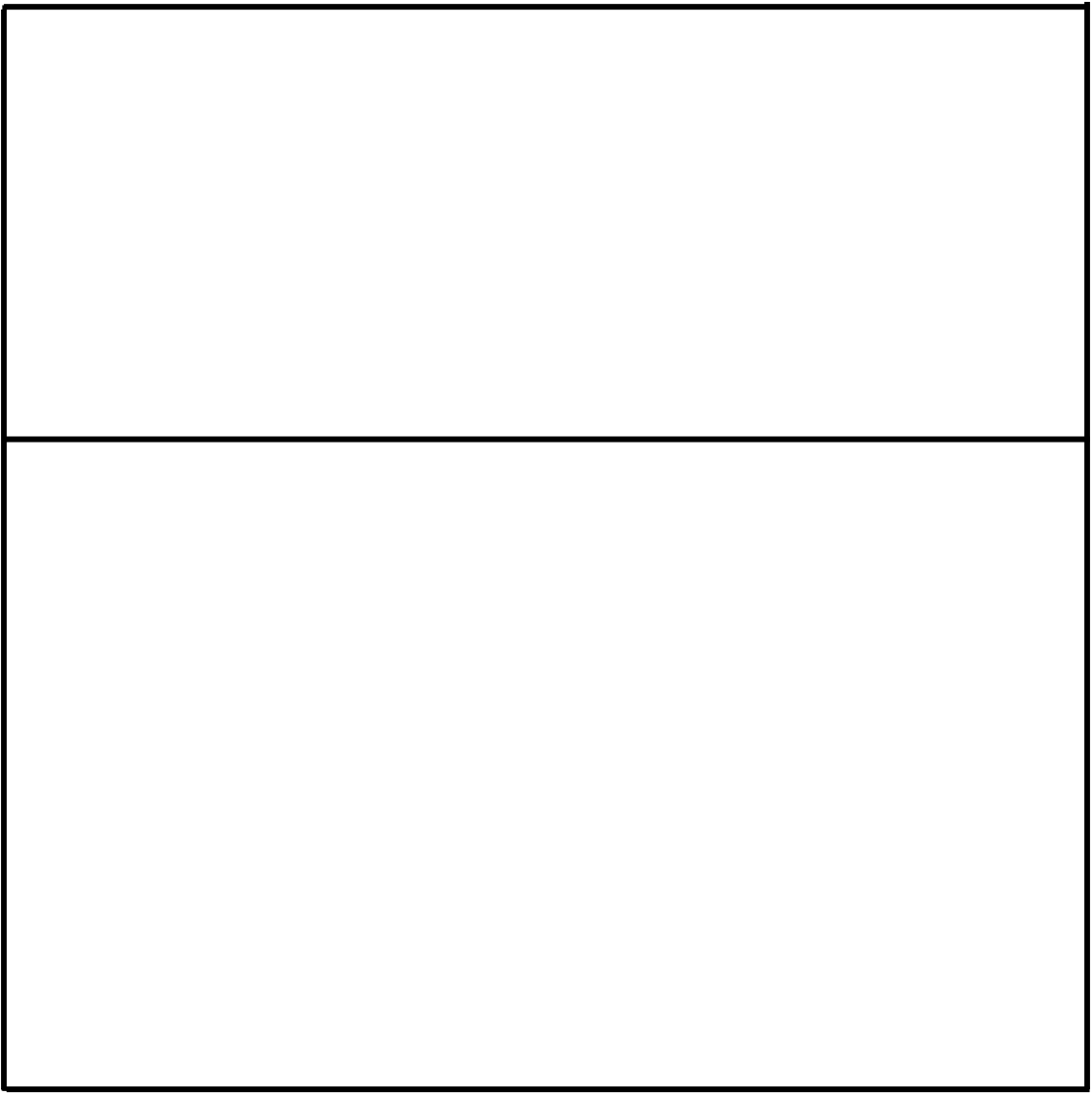
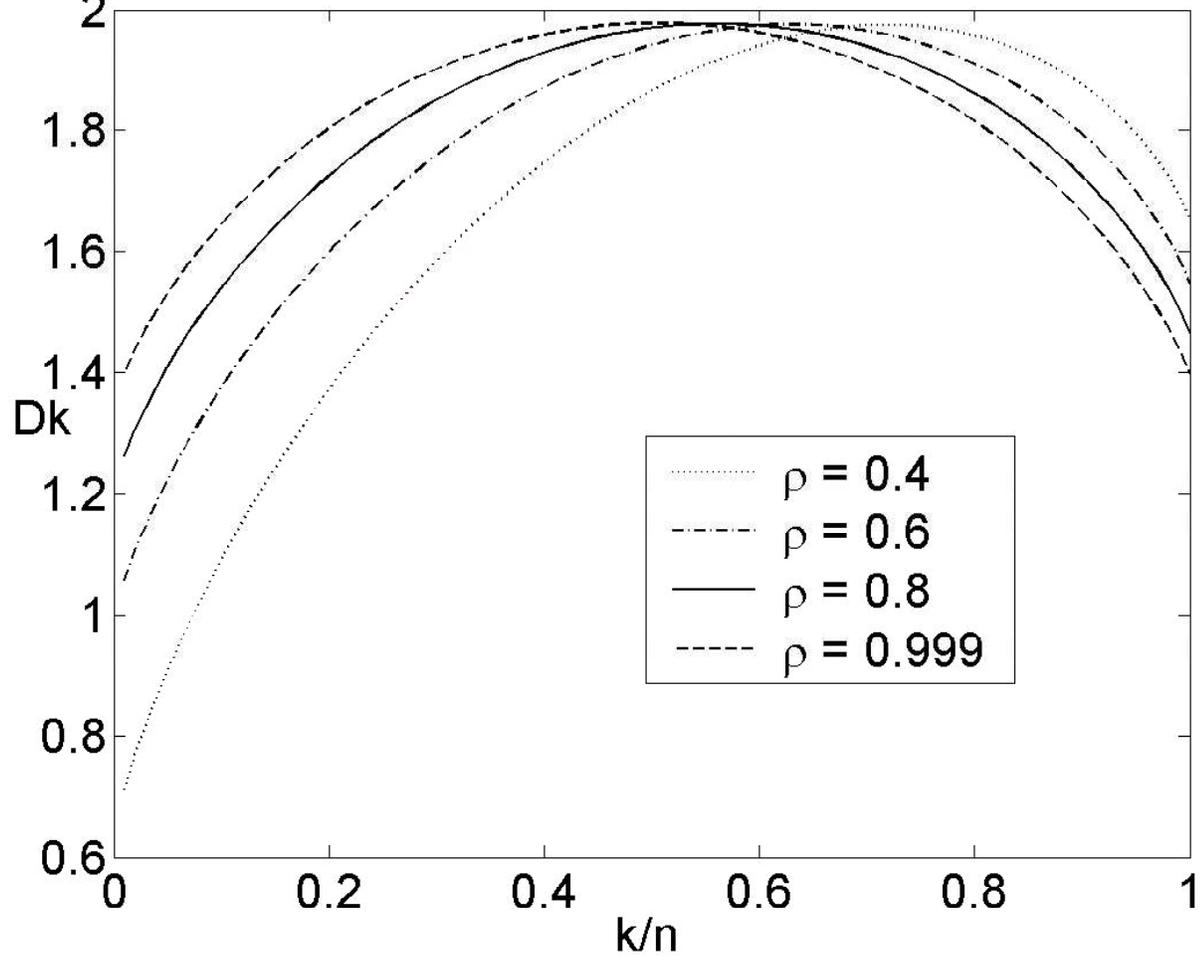
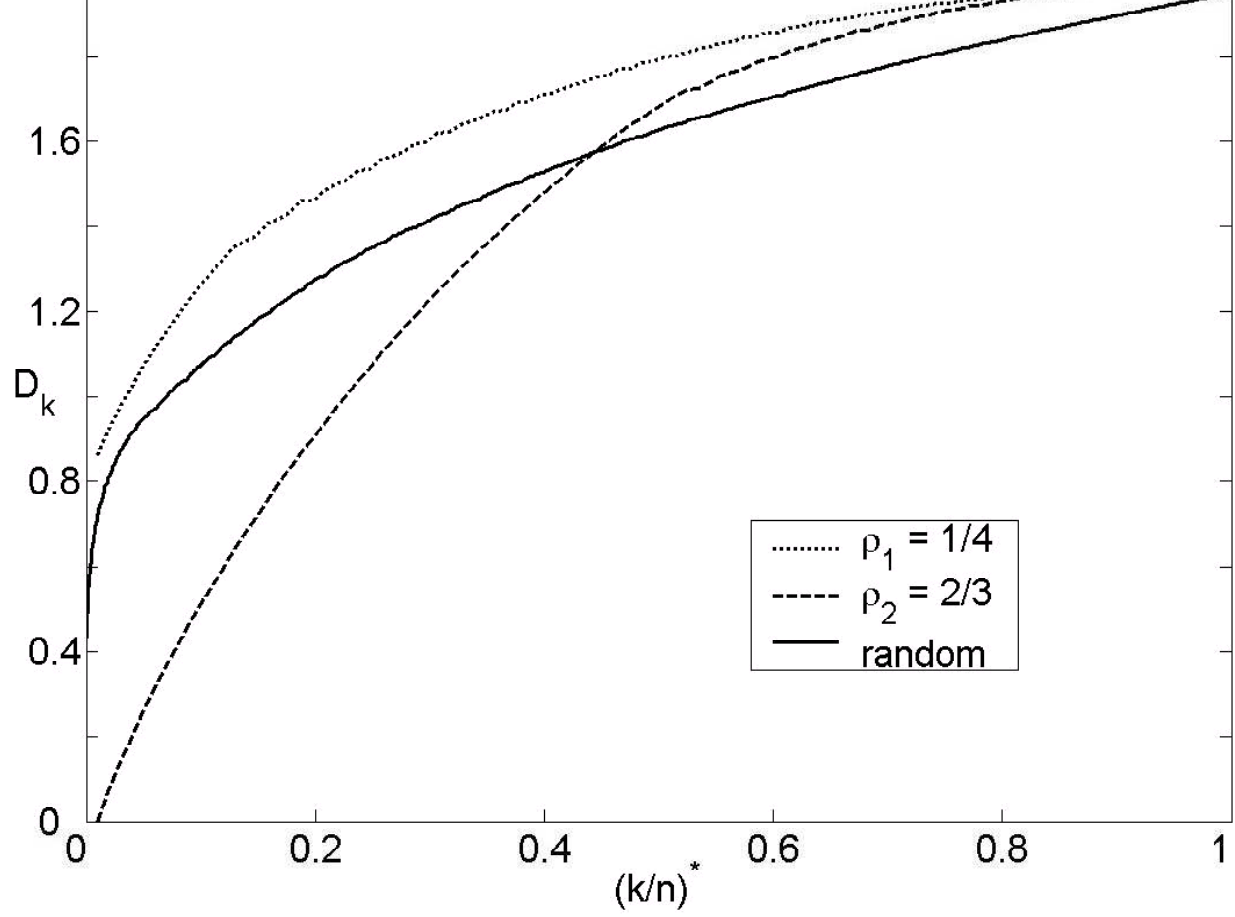
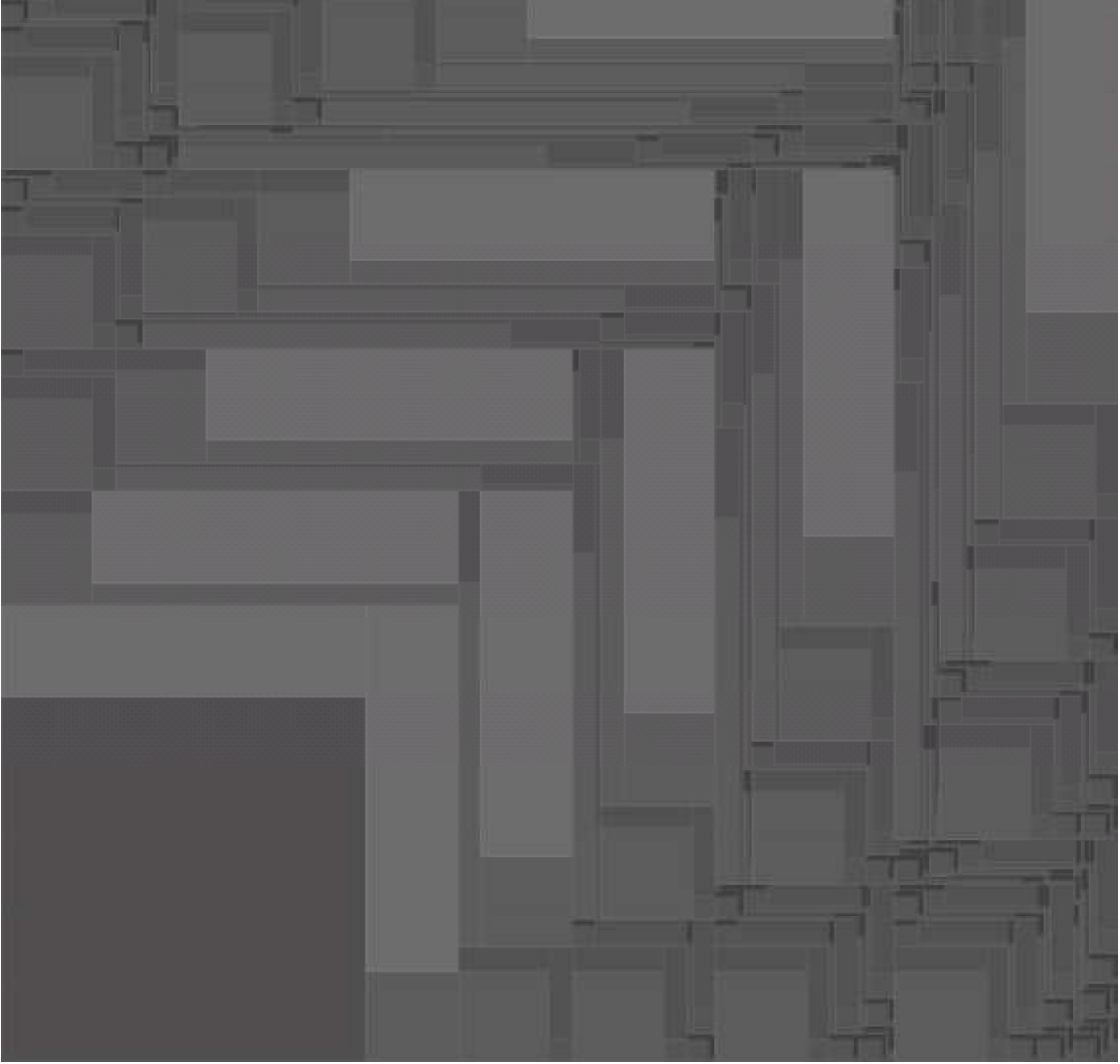
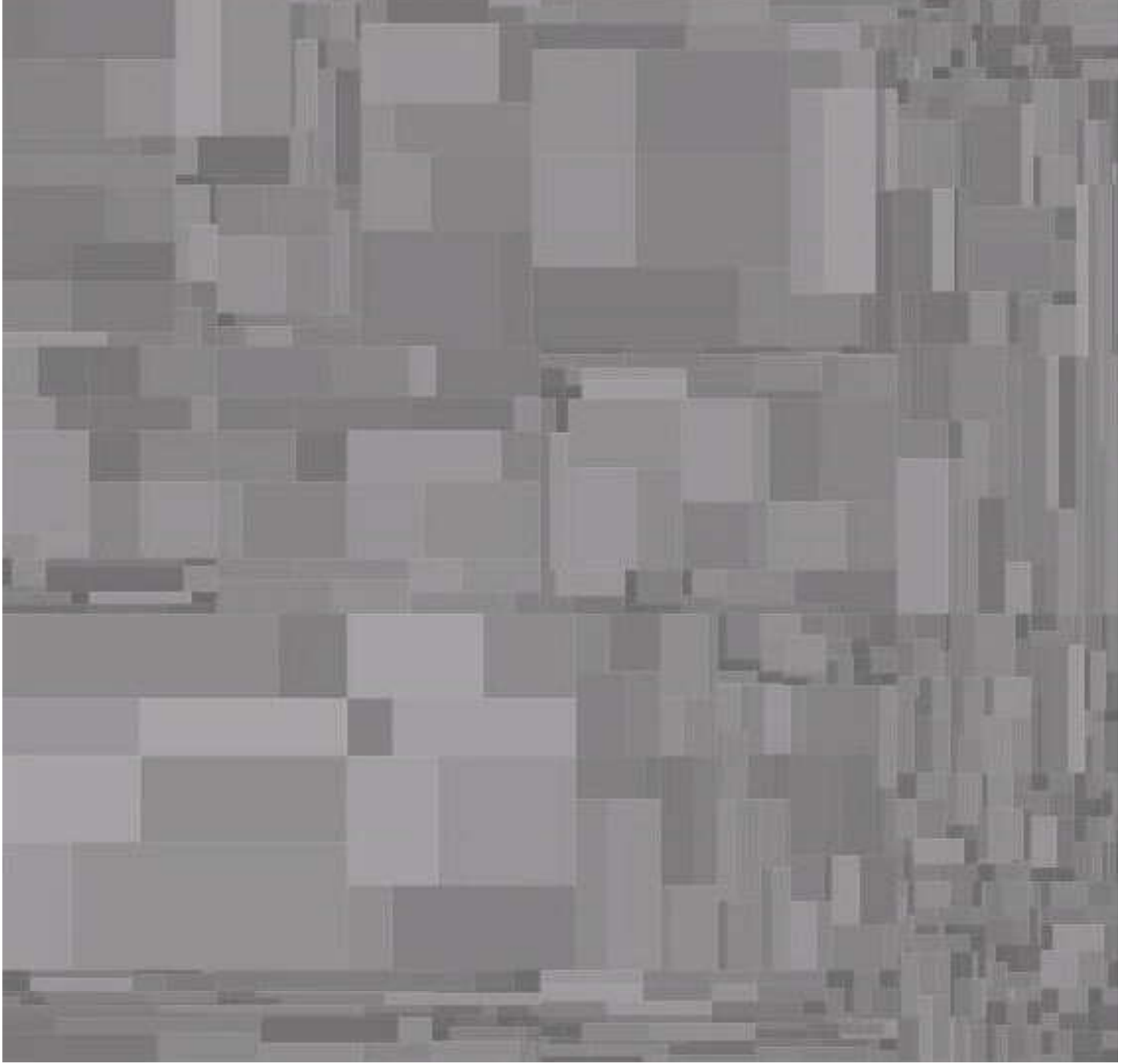
\end{document}